\documentstyle[aps,prl,psfig,multicol]{revtex}    

\begin{document}

\title{\bf Extreme Value Statistics and Traveling Fronts: Various Applications}

\author{Satya N. Majumdar$^{1}$ and P. L. Krapivsky$^{2}$}
\address{
 {\small $^1$Laboratoire de Physique Quantique (UMR C5626 du CNRS),
Universit\'e Paul Sabatier, 31062 Toulouse Cedex, France}\\
{\small $^2$Center for Polymer Studies and
Department of Physics, Boston University, Boston, MA 02215, USA}}

\maketitle
\widetext   

\begin{abstract}

\noindent
An intriguing connection between extreme value statistics and traveling
fronts has been found recently in a number of diverse problems.  In this
short review we outline a few such problems and consider their various
applications.

\end{abstract}

\bigskip

\begin{multicols}{2}

Independence of random events is the most desirable property in probability
theory and statistical physics. If we have a collection of independent random
variables $X_1,\ldots,X_N$ with finite variance, then the distribution of the
sum ${\sum X_j}/{\sqrt{N}}$ is Gaussian in the thermodynamic limit $N\to \infty$.
Similarly under broad circumstances, the asymptotic distribution of the
extreme values, e.g., $X_{\rm min}={\rm min}(X_1,\ldots,X_N)$, belongs to one
of just three possible families\cite{F,G,Gumbel}.
  
However, independence is the exception rather than the rule -- random
variables are often highly correlated.  Little is known on extreme value
statistics of correlated random variables yet a vast number of problems can
be recast into such scheme.  The celebrated example is the traveling salesman
problem, that is to find the shortest closed tour visiting every `city' once.
There are $(N-1)!/2$ possible tours and the lengths of the tours are
obviously correlated. This and a few other combinatorial optimization
problems were recently analyzed by using techniques originally developed to
study spin glasses\cite{MPV}.  A nice general review of the recent progress
in that direction is given by Martin, Monasson and Zeccina\cite{MMZ}; for an
outstanding progress in one specific problem, the matching problem,
see\cite{MPA}.

In a series of recent publications\cite{KM,MK,MKE,DM,BKM,MKC} we have shown
that there is an intriguing connection between the statistics of extreme
values arising in various contexts and traveling fronts. More precisely, the
cumulative distributions of extreme variables were shown to admit a traveling
front solution.  Such a connection was also noted in the context of a
particle moving in a random potential\cite{C+D}. The goal of this short
review is to convince the readers that the techniques of traveling fronts is
a powerful tool to tackle the extreme value statistics of correlated random
variables.  By casting the problem in the traveling front framework, one
easily determines two leading terms in the asymptotic expansion of the
average value of the extreme variable.  Furthermore, the variance of the
extreme variable is nothing but the width of the traveling wave front and
therefore it is usually {\em finite}.  These results are very natural in the
traveling wave framework yet very difficult to guess and derive using other
methods.

Traveling front solutions have been found in numerous 
problems\cite{van2}.  To keep the discussion short, we consider the most well
known example -- the one-dimensional KPP (Kolmogorov, Petrovsky, Piskunov)
equation\cite{KPP}, also known as the Fisher equation\cite{Fisher}. This
is a nonlinear partial differential equation
\begin{equation}
{\partial \phi\over \partial t}
={\partial^2 \phi\over \partial x^2}+\phi-\phi^2,
\label{kpp}
\end{equation} 
where $\phi(x,t)$ represents, for example, the density of a population at a
point $x$ at time $t$. Clearly this equation has two fixed points or
stationary solutions: (i) $\phi (x)=1$ for all $x$ and (ii) $\phi(x)=0$ for
all $x$. A simple linear stability analysis shows that the solution (i) is
stable while the solution (ii) is unstable.  Therefore, if one starts with a
sufficiently sharp initial condition, say $\phi(x,t=0)=1$ for $x<0$ and
$\phi(x,t=0)=0$ for $x\ge 0$ it is easy to see (for example by numerical
simulation) that as time proceeds, the front separating the stable solution
$\phi=1$ and the unstable solution $\phi=0$ advances in the forward direction
with a unique velocity $v_f$. Besides, the front retains its shape in the
sense that the width of the front remains finite even at large times. The
front velocity $v_f$ is determined by analyzing the tail region $x\to
\infty$. In this region, $\phi$ is small and one can ignore the nonlinear
term $\phi^2$ in Eq.~(\ref{kpp}).  The resulting linear equation allows a
spectrum of decaying solutions $\phi(x,t)\propto e^{-\lambda[x-v(\lambda)t]}$
provided $v(\lambda)$ satisfies the dispersion relation
\begin{equation}
v(\lambda)= \lambda+ {1\over {\lambda}}.
\label{dispersion}   
\end{equation}
Thus a whole family of solutions parametrized by $\lambda$ is in principle
allowed.  However, the front actually advances with a unique velocity $v_f$.
Thus there must be a selection principle to choose the right velocity from
the whole spectrum $v(\lambda)$.  Note that the dispersion spectrum 
(\ref{dispersion}) has a unique minimum at $\lambda=\lambda^*=1$ where
$v(\lambda^*)=2$. It was shown\cite{KPP,front1} that for sufficiently steep
initial conditions, the extremum of the dispersion curve is selected by the
front, i.e., $v_f =v(\lambda^*)=2$.  Note that while the spectrum is
determined solely by the linearized equation, for a given initial condition
the nonlinear term plays a crucial role in selecting the final velocity from
the full spectrum allowed by the linear equation. Subsequently it was
shown\cite{front1,van1,B+D} that the front position $x_f(t)$, apart from the
leading $v(\lambda^*) t$ term, has a slow logarithmic correction
\begin{equation}
x_f(t) = v(\lambda^*)t - {3\over {2\lambda^*}}\ln t + \ldots
\label{correction}
\end{equation}   

Although this velocity selection principle was originally
proved only for the KPP equation, this strategy of selecting the extremum of
the dispersion spectrum of the linearized equation was subsequently shown to
apply to various traveling front solutions provided certain conditions are
satisfied\cite{van2}.  We will show how this
selection principle can be successfully used to derive exact asymptotic
results for the statistics of extreme variables in a number of problems. 
In all the problems discussed below, we will
find a traveling front solution of the same generic form (with a leading
`linear' term followed by a subleading `logarithmic' correction) as in Eq.
(\ref{correction}).  While the the velocity dispersion spectrum $v(\lambda)$
will be widely different from problem to problem, the principle of selecting
the extremum of the spectrum, namely $v(\lambda^*)$, will still be valid.
Finally, while $\lambda^*$ and $v(\lambda^*)$ are thus nonuniversal, the
prefactor $3/2$ in the logarithmic correction term turns out to be universal
and is just the first excited state energy of a quantum harmonic
oscillator\cite{B+D,KM,MKC}. For a short derivation of this correction term,
see e.g. appendix A of Ref.\cite{MKC}.

\vspace{0.2cm}

\noindent {\bf {Directed Polymer on a Cayley Tree:}} 
As a first example, consider the problem of directed polymer on a Cayley tree
studied by Derrida and Spohn\cite{D+S} and recently resurfaced in a number of
apparently unrelated problems\cite{C+D,CdC,tang,BM}.  The primary emphasis of
this work was on the spin glass like transition occurring at {\em finite}
temperature and on {\em fluctuation} properties.  In contrast, we consider
exactly {\em zero} temperature and focus on the basic {\em macroscopic}
quantity, namely the ground state energy\cite{MK,DM}.

We consider a tree rooted at $O$ (see Fig. 1) where a random energy
$\epsilon$ is associated with every bond of the tree. The variables
$\epsilon$'s are independent and each drawn from the same distribution
$\rho(\epsilon)$. A directed polymer of size $n$ goes down from the root $O$
to any of the $2^n$ nodes at the level $n$. There are $N=2^n$ possible
paths for the polymer of size $n$ and the energy of any of these paths is
\begin{equation}
E_{\rm path}=\sum_{i\in {\rm path}}{\epsilon}_i .
\label{energy}
\end{equation}
The set of $N=2^n$ variables $E_1$, $E_2$, $\ldots$, $E_N$ are clearly
correlated in a hierarchical (i.e., ultrametric) way and the two point
correlation between the energies of any two paths is proportional to the
number of bonds they share.  The ground state energy $E_{\rm min}(n)={\rm
  min}[E_1,E_2,\ldots, E_{2^n}]$ is then a random variable and we are
interested in its statistics.
\begin{figure}
\begin{center}
\leavevmode
\psfig{figure=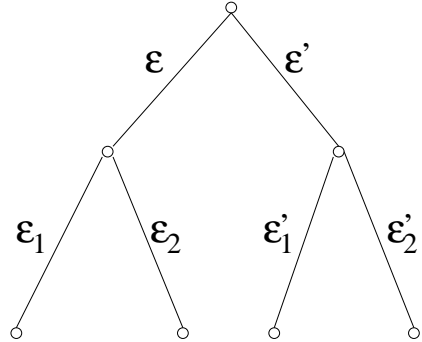,width=8cm,angle=0}
 \caption{ The directed polymer on a Cayley tree. The $\epsilon$'s denote
the bond energies.}
\end{center}
\end{figure} 

The suitable quantity that has the traveling front solution is the cumulative
distribution $P_n(x)={\rm Prob}[E_{\rm min}\ge x]$.  It satisfies\cite{MK} a
nonlinear recursion relation,
\begin{equation}
P_{n+1}(x)={\left[ \int d\epsilon \rho(\epsilon) P_n(x-\epsilon)\right]}^2,
\label{recur1}
\end{equation}
and the initial condition $P_0(x)=\theta(-x)$, where $\theta(x)$ is
the Heaviside step function. The front velocity can be determined by
following the `linearizing' strategy outlined above for the KPP equation.
Unlike the KPP equation, however, we now need to linearize the recursion
equation (\ref{recur1}) in the `backward' tail region $x\to
-\infty$\cite{MK}.  By inserting $1-P_n(x)\propto
e^{-\lambda[x-v(\lambda)n]}$ into Eq.~(\ref{recur1}), one finds\cite{MK} a
dispersion spectrum
\begin{equation}
v(\lambda)= 
-{1\over {\lambda}}\ln {\left[ 2 \int d\epsilon \rho(\epsilon) 
e^{-\lambda \epsilon}\right]}.
\label{dispersion1}
\end{equation}
For generic distributions $\rho(\epsilon)$, this spectrum has a unique
maximum at $\lambda=\lambda^*$ and by the general velocity selection
principle the maximum velocity $v(\lambda^*)$ will be selected by the front.
We now illustrate it for two distributions.

\vspace{0.2cm}

\noindent {\bf {(i)}} The bimodal distribution,
  $\rho(\epsilon)=p\delta(\epsilon-1) +(1-p)\delta(\epsilon)$, where $0\le
  p\le 1$. In this case one can also think of the energy of a bond as a
  `length' variable and a bond is present with probability $p$ and absent
  with probability $(1-p)$. By analysing Eqs. (\ref{recur1}) and
  (\ref{dispersion1}) it was found\cite{MK} that the polymer undergoes a
  `depinning' transition at the critical value $p_c=1/2$. For $p>1/2$, the
  ground state energy is `extensive' and increases linearly with $n$. On the
  other hand, for $p<1/2$ the polymer is `localized' and the ground state
  energy remains finite even in the $n\to \infty$ limit. More precisely, the
  average ground state energy $\langle E_{min}(n)\rangle = \int_0^{\infty}
  P_n(x)dx$ has the asymptotic behaviors\cite{MK}
\begin{equation}
\label{Ln}
\langle E_{\rm min}(n)\rangle\simeq\cases
                    {v_{\rm min}(p)n                 &$p>1/2$,\cr
                              (\ln 2)^{-1}\ln\ln n   &$p=1/2$,\cr
                              {\rm finite}           &$p<1/2$,\cr}
\end{equation} 
where $v_{\rm min}(p)= v(\lambda^*)$ with $v(\lambda^*)$ being the maximum of
the dispersion spectrum in Eq.~(\ref{dispersion1}). Taking into account the
correction term as in the KPP equation, we found that for $p>1/2$, the
average ground state energy is given by\cite{MK}
\begin{equation}
\langle E_{\rm min}(n)\rangle
\simeq v_{\rm min}(p)\,n+{3\over 2\lambda^*}\,\ln n +\cdots
\label{asymp}
\end{equation} 
In a similar way, one can compute the average maximum energy.  $\langle
E_{\rm max}(n)\rangle \simeq v_{\rm max}(p)\,n$.  Interestingly, the
velocities satisfy a duality relation $v_{\rm min}(p)+v_{\rm max}(1-p)=1$.

\vskip 0.2cm

\noindent {\bf  {(ii)}} For the unbounded one sided distribution 
$\rho(\epsilon)=e^{-\epsilon}\theta(\epsilon)$, the dispersion spectrum
(\ref{dispersion1}) becomes 
\begin{equation}
v(\lambda)= -{1\over \lambda}
\ln\left[{2\over \lambda+1}\right].  
\label{dispersion2}
\end{equation} 
The average ground state energy is given by Eq.~(\ref{asymp}) with
$\lambda^*=3.31107 \ldots$ and $v(\lambda^*)=0.23196 \ldots$.

We have also studied the asymptotic behaviors of the full distribution of the
ground state energy for various $\rho(\epsilon)$ and found\cite{DM} that the
hierarchical correlations between the random variables $[E_1,E_2,\ldots,
E_{2^n}]$ violate the well known Gumbel type behaviors\cite{Gumbel} exhibited
by the the distribution of the extreme of a set of {\em independent} random
variables.

\vskip 0.2cm

\noindent {\bf Random Binary Search Tree:} 
We now outline an application of the traveling front techniques to analysis
of search algorithms in computer science\cite{MKC}.  A computer is constantly
fed with enormous amount of data. It is therefore essential to organize or
sort the data in an efficient way so that the computer spends the minimum
time to search for a data if required later. There are various `sorting' and
`search' algorithms devised for this purpose\cite{knuth}. One particular
algorithm that has been widely studied by computer scientists is the Random
Binary Search Tree (RBST) algorithm\cite{mahm}.  To understand this
algorithm, consider a simple example. Suppose the incoming data string
consists of the twelve months of the year appearing in the following random
order: July, September, December, May, April, February, January, October,
November, March, June and August. 
\begin{figure}
\begin{center}
\leavevmode
\psfig{figure=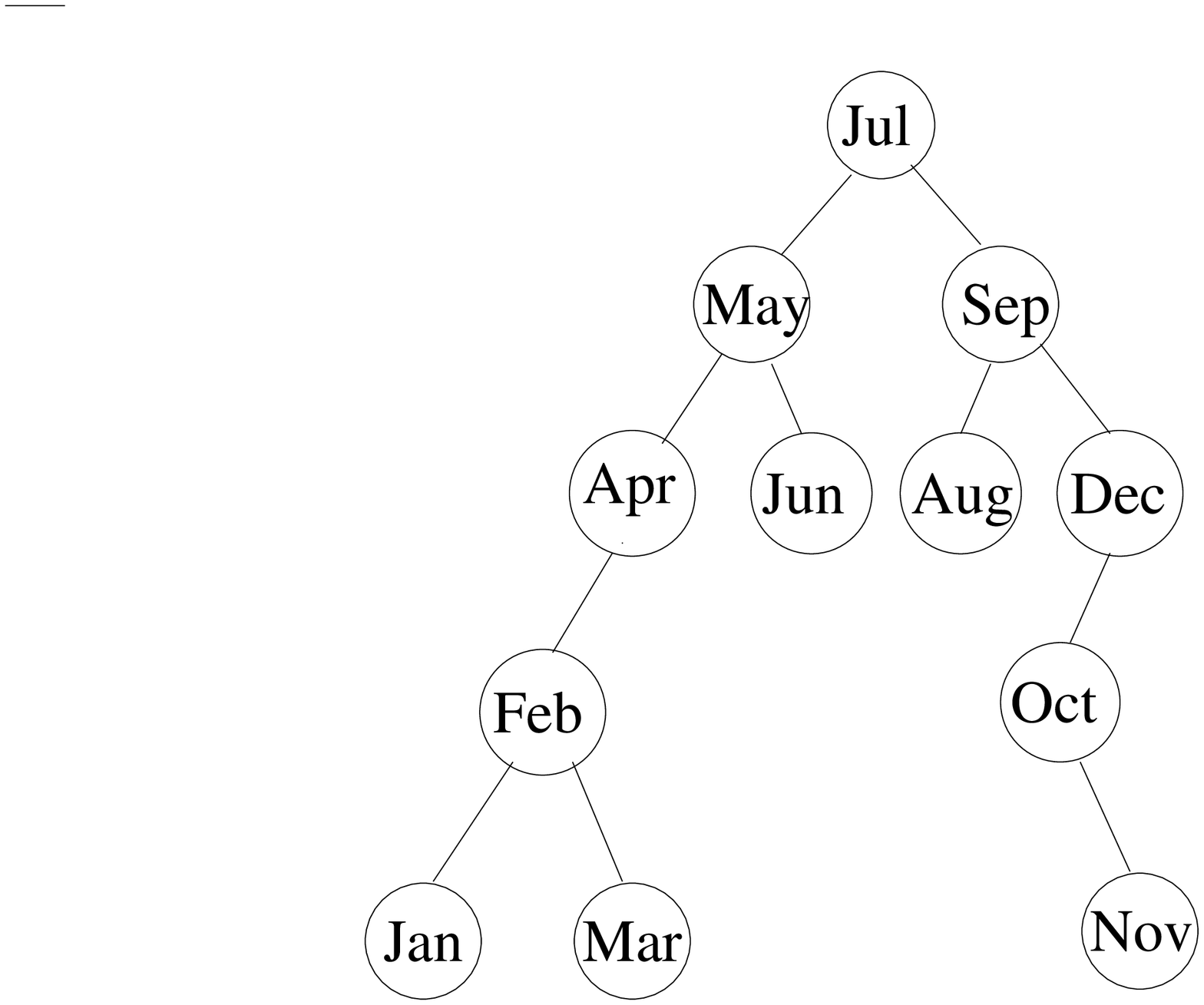,width=8cm,angle=0}
 \caption{The binary search tree corresponding
to the data string in the order: July, September, May, April, February,
January, October, November, March, June, and August.
The tree has size $N=12$ and height $H_N=5$.}
 \end{center}
\end{figure}

The RBST algorithm stores this data on a
binary tree in the following way. A chronological order (January, February,...)
is first chosen. Now the first element of the input string (July) is put at
the root of a tree (see Fig. 2).  The next element of the string is
September. One compares with the root element (July) and sees that September
is bigger than July (in chronological order). So one assigns September to a
daughter node of the root in the right branch. On the other hand, if the new
element were less than the root, it would have gone to the daughter node of
the left branch. Then the next element is December. We compare at the root
(July) and decide that it has to go to the right, then we compare with the
existing right daughter node (September) and decide that December has to go
to the node which is the right daughter of September. The process continues
till all the elements are assigned their nodes on the tree.  For the
particular data string in the above example, we get the unique tree shown in
Fig.~2.  If the incoming data (consisting of $N$ elements) is completely
random, then all $N!$ possible binary trees have equal probability to
occur and are called random binary search trees.
            
A quantity that is widely used to measure the efficiency of such an algorithm
is the maximum search time required to find an element. This is the worst
case scenario.  The maximum search time is quantified by the height $H_N$ of
a tree, i.e., the distance from the node to the farthest element on the tree.
In the example of Fig.~2, $H_N=5$.  Thus determining the statistics of $H_N$
is an extreme value problem.  Apart from a slight modification, this problem
can be mapped\cite{MKC} onto the directed polymer problem described in the
previous example. In the modified problem, the bond energies are not
completely uncorrelated as in the usual polymer problem, but are correlated
in a special way: The energies of two bonds emanating from the same parent
(see Fig.~1) satisfy the constraint, $e^{-\epsilon} +e^{-\epsilon'}=1$ and
this constraint holds at every level of the tree. For example, in Fig.~1, we
also have $e^{-\epsilon_1}+e^{-\epsilon_2}=1$,
$e^{-\epsilon_1'}+e^{-\epsilon_2'}=1$, etc.  The statistics of the height
variable $H_N$ in RBST problem was shown to be identical to the ground state
energy $E_{\rm min}(n)$ of this modified polymer problem with the
identification $N=2^n$.  The traveling front analysis then gives the average
height\cite{MKC}
\begin{equation}
\langle H_N\rangle \simeq \alpha_0 \ln N +
\alpha_1 \ln \ln N.
\label{HN1}
\end{equation}
Here $\alpha_0 = 1/v(\lambda^*)$ and $\alpha_1= - 3/{2\lambda^*
  v(\lambda^*)}$ where $\lambda^*=3.31107 \ldots$ is the maximum of the
dispersion curve (\ref{dispersion2}) with $v(\lambda^*)=0.23196
\ldots$.  The traveling fronts retains its shape asymptotically,
i.e., its width remains finite. This shows that the variance of the height
$H_N$ remains finite even in the large $N$ limit.  While
some of these results mentioned here were also derived by the computer
scientists using rigorous mathematical
bounds\cite{Devroye,Reed,Drmota}, the existence of a traveling front
was not realized before. Besides, the traveling front techniques allow us to
derive more detailed results such as the asymptotic behaviors of the full
probability distribution of $H_N$ (not just its moments) and also obtain
completely new results for trees generated with arbitrary distributions, not
necessarily uniform\cite{MKC}. The statistics of other observables such as
the `balanced' height of a tree (which corresponds to the maximum energy in
the modified directed polymer problem) has also been derived exactly using
the traveling front techniques\cite{MKC}.

\vskip 0.2cm

\noindent {\bf Aggregation Dynamics of Growing Random Trees:} 
In the RBST example discussed above, the trees have fixed size $N$.
Alternately one can consider random trees generated dynamically via an
aggregation mechanism where the size of the trees grows indefinitely with
time $t$. Apart from the computer science problems discussed above, such
growing trees also arise in physical situations such as collision processes
in gases\cite{van} where the largest Lyapunov exponent is related to the
maximum height problem.

We studied a simple tree generation model\cite{BKM} where initially we have
an infinite number of trivial (single-leaf) trees.  Then, two trees are
picked at random and attached to a common root.  This merging process is
repeated indefinitely (to simplify formulas the rate is set equal to 2).  Let
$c(t)$ be the number density of trees at time $t$.  Initially, $c(0)=1$, and
then it evolves according to $dc/dt=-c^2$ whose solution is
$c(t)=(1+t)^{-1}$.  Thence the average number of leaves per tree $\langle
N\rangle $ grows linearly with time, $\langle N\rangle =c^{-1}=1+t$. We are
interested in the minimal and maximal heights of such a growing tree.  Rather
than considering the two extremal height distributions separately, we studied
a more general model that interpolates between the two cases.  In this model,
each tree carries an extremal height $k$.  The result of a merger between
trees with extremal heights $k_1$ and $k_2$ is a new tree with extremal
height $k$ given by
\begin{equation}
\label{min}
k=\cases{ {\rm min}(k_1, k_2)+1 & with prob.~ $p$, \cr
{\rm max}(k_1,k_2)+1 & with prob.~ $1-p$ .\cr}
\end{equation}
Here, $p$ is a mixing parameter whose limits $p=1$ and $p=0$ correspond
to the minimal and the maximal heights problems, respectively.   

The number density of trees with extremal height $k$, $c_k(t)$,
evolves according to the master equation
\begin{equation}
\label{ckmin}
{d c_k\over dt}=c_{k-1}^2\!\!-2cc_k\!+\!2pc_{k-1}
\!\!\sum_{j=k}^\infty c_j\!+\!2(1\!-\!p)c_{k-1}\!\!\sum_{j=0}^{k-2}c_j.
\end{equation}
Introducing the cumulative fractions
$A_k=c^{-1}\sum_{j=k}^\infty c_j$
and a new time variable 
$T=\int_0^t d\tau\,c(\tau)=\ln(1+t)$,
we recast Eqs.~(\ref{ckmin}) into
\begin{equation}
\label{Akmin}
{dA_k\over dT}=-A_k+2(1-p)A_{k-1}+(2p-1)A_{k-1}^2,
\end{equation}
which should be solved subject to the step function initial
conditions, $A_k(0)=1$ for $k\leq 0$ and $A_k(0)=0$ otherwise.

In the long time limit, $A_k(T)$ approaches a traveling wave form, $A_k(T)\to
f(k-vT)$.  The velocity $v$ can be determined\cite{BKM} as in the above
problems.  Knowing this velocity one can compute the expected extremal tree
height from the relation, $\langle k\rangle=c^{-1}\sum_k kc_k$. Let us note
one interesting fact.  For $p=0$ (the maximal height case), we can express
$\langle k\rangle $ as a function of $\langle N\rangle$ (after eliminating
$t$ using $\langle N\rangle=1+t$) for large $\langle N\rangle$ and find 
\begin{equation}
\langle k\rangle= \alpha_0 \ln{\langle N \rangle}
+ \alpha_1\,\ln \ln {\langle N\rangle},
\label{xmin}
\end{equation}
where $\alpha_0$ and $\alpha_1$ identical to those in Eq.~(\ref{HN1}) for the
fixed $N$ trees. This shows that the dynamically growing trees have the same
asymptotic properties as those of the fixed size trees for large trees if one
replaces $N$ in Eq.~(\ref{HN1}) by $\langle N\rangle$.  This dynamic approach
is thus similar to the grand canonical approach in statistical mechanics with
the time $t$ playing the role of the chemical potential that can be chosen to
fix the average size.

\vskip 0.2cm

\noindent {\bf The Dynamics of Efficiency in a Simple Model:} 
We now move from trees to economy and consider a model that mimics the
dynamics of efficiencies of competing agents\cite{MKE}. We represent the
efficiency of each agent by a single nonnegative number.  The efficiency of
every agent can, independent of other agents, increase or decrease
stochastically by a certain amount which we set equal to unity.  In addition,
the agents interact with each other which is the fundamental driving
mechanism for economy.  We assume that the interaction equates the
efficiencies of underachievers to the efficiencies of better performing
agents and set the rate of this process equal to one; we denote the rates of
the increase and decrease of the efficiency by $p$ and $q$ respectively.

Let $h_i(t)$ is the efficiency of the agent $i$ at time $t$ and $P(h,t)$ is
the fraction of agents with efficiency $h$ at time $t$.  The evolution
equation for $P(h,t)$ is obtained by counting all possible gain and loss
terms. For $h\ge 1$, this equation reads\cite{MKE}
\begin{eqnarray}
{dP(h,t)\over dt}&=&-P(h,t)
\sum_{h'=h+1}^{\infty}P(h',t)-(p+q)P(h,t) \nonumber \\
&+& q P(h+1,t) +pP(h-1,t) \nonumber \\
&+& P(h,t)\sum_{h'=0}^{h-1} P(h',t).
\label{pht}
\end{eqnarray}
The cumulative distribution $F(h,t)=\sum_{h'\geq h} P(h',t)$ satisfies
\begin{eqnarray}
{dF(h,t)\over dt}&=&-F^2(h,t)+(1-p-q)F(h,t) \nonumber\\ 
&+&q F(h+1,t)+pF(h-1,t).
\label{fht}
\end{eqnarray}  
The function $F(h,t)$ approaches a traveling wave form,
$F(h,t)=f(h-vt)$. The velocity can
be determined\cite{MKE} by repeating the steps detailed in the previous
examples.  The analysis is more cumbersome due to appearance of the
critical line $p_c(q)$ in the $(p,q)$ plane,
\begin{equation}
\label{critic}
p_c(q)=\cases{1+q-2\sqrt{q} & for $q\ge 1$, \cr
              0  & for $q\le 1,$}
\end{equation}         
separating different behaviors.  For $p>p_c(q)$, the system is in the
developing phase with the average efficiency $\langle h\rangle$ increasing
according to
\begin{equation}
\langle h\rangle=v_{\rm min}t-{3\over 2\lambda^*}\,\ln t
+{\cal O}(1),
\label{avh}
\end{equation}
Here again $\lambda^*$ is the decay rate, $f(x)\propto e^{-\lambda^*x}$ as
$x\to\infty$.  For $p\le p_c(q)$ with $q>1$, the system is localized and
$\langle h\rangle$ approaches a time-independent constant in the long time
limit.  For $p=0$ and $q<1$, the system is in the developing phase for
unbounded initial efficiency distributions, with the growth rate dependent on
initial conditions.  For economically relevant compact initial conditions,
the regime $p=0$ and $q<1$ belongs to the stagnant phase.

\medskip
To summarize, we have exemplified that the asymptotic statistics of extreme
values can often be analytically determined using the powerful traveling
front techniques.  There might be a deeper hidden connection between the
extreme value statistics and traveling fronts, and establishing such a
connection is a challenging task.  Finally we mention that the recently
computed\cite{KaulP} correction to the the celebrated Bekenstein-Hawking (BH)
formula for the black hole entropy $S_{\rm BH}=A_{\rm H}/4l_{\rm Pl}^2$ (here
$A_H$ is the classical horizon area and $l_{\rm Pl}$ is the Planck length) is
logarithmic. Specifically, for the four-dimensional non-rotating black hole,
the entropy expansion\cite{KaulP}
\begin{equation}
S=S_{\rm BH} - {3\over {2}}\,\ln S_{\rm BH}  + \ldots
\label{bh}
\end{equation} 
strikingly resembles the time dependence of the front position, 
Eq.~(\ref{correction}). The elogarithmic correction term 
also appears for other black holes\cite{DMB}.  It is therefore tempting to
speculate a traveling front structure in the black hole entropy. This
speculation gets further strengthened by the fact that the BH entropy is the
`maximal' entropy that a black hole can have\cite{KaulP}.  In other words
this may be considered as an extreme value problem.  The verification of the
existence, if any, of a traveling front structure in the black hole problem
remains an outstanding open problem.

\medskip
We thank E.~Ben-Naim and D.~S.~Dean for collaborations on some of these
topics. We also thank J.~K.~Bhattacharjee for pointing out the possible
connection to the black hole entropy problem.

\end{multicols}

\end{document}